\documentclass[manuscript, nonacm]{acmart}
\usepackage{listings}

\usepackage{colortbl}

\usepackage{xcolor}
\usepackage{todonotes}

\definecolor{codegreen}{rgb}{0,0.6,0}
\definecolor{codegray}{rgb}{0.5,0.5,0.5}
\definecolor{codepurple}{rgb}{0.58,0,0.82}
\definecolor{backcolour}{rgb}{0.95,0.95,0.92}

\lstdefinestyle{mystyle}{
    backgroundcolor=\color{backcolour},   
    commentstyle=\color{codegreen},
    keywordstyle=\color{magenta},
    numberstyle=\tiny\color{codegray},
    stringstyle=\color{codepurple},
    basicstyle=\ttfamily\footnotesize,
    breakatwhitespace=false,         
    breaklines=true,                 
    captionpos=b,                    
    keepspaces=true,                 
    numbers=left,                    
    numbersep=5pt,                  
    showspaces=false,                
    showstringspaces=false,
    showtabs=false,                  
    tabsize=2
}

\newcommand{\rqOne}[0]{How useful are the LLM-powered chatbot and the search functionality, and which functionality do the participants prefer?}

\newcommand{\rqTwo}[0]{How do the LLM-powered chatbot and the search functionality relate to the participant performance in the information-seeking tasks?}

\usepackage{balance}
\usepackage{multirow}

\AtBeginDocument{%
  }

\setcopyright{acmlicensed}
\begin{document}

\title{Comparing the Utility, Preference, and Performance of Course Material Search Functionality and Retrieval-Augmented Generation Large Language Model (RAG-LLM) AI Chatbots in Information-Seeking Tasks}

\author{Leonardo Pasquarelli}
\affiliation{%
  \institution{Aalto University}
  \city{Espoo}
  \country{Finland}}
\email{pasquarelli.le@gmail.com}

\author{Charles Koutcheme}
\affiliation{%
  \institution{Aalto University}
  \city{Espoo}
  \country{Finland}
}
\email{charles.koutcheme@aalto.fi}

\author{Arto Hellas}
\affiliation{%
  \institution{Aalto University}
  \city{Espoo}
  \country{Finland}
}
\email{arto.hellas@aalto.fi}

\begin{abstract}
Providing sufficient support for students requires substantial resources, especially considering the growing enrollment numbers. Students need help in a variety of tasks, ranging from information-seeking to requiring support with course assignments. To explore the utility of recent large language models (LLMs) as a support mechanism, we developed an LLM-powered AI chatbot that augments the answers that are produced with information from the course materials. To study the effect of the LLM-powered AI chatbot, we conducted a lab-based user study (N=14), in which the participants worked on tasks from a web software development course. The participants were divided into two groups, where one of the groups first had access to the chatbot and then to a more traditional search functionality, while another group started with the search functionality and was then given the chatbot. We assessed the participants' performance and perceptions towards the chatbot and the search functionality and explored their preferences towards the support functionalities. Our findings highlight that both support mechanisms are seen as useful and that support mechanisms work well for specific tasks, while less so for other tasks. We also observe that students tended to prefer the second support mechanism more, where students who were first given the chatbot tended to prefer the search functionality and vice versa.
\end{abstract}


\maketitle

\section{Introduction}

The recent advances in large language models (LLMs) have surprised many computing educators and researchers, leading to explorations of what is possible with LLMs and what the broader effects of LLMs are on computing education~\cite{prather2023robots,denny2024computing}. LLMs have been shown to have potential, especially in programming and programming education, where LLMs can be used to create and solve programming problems~\cite{sarsa2022automatic,finnie2023my,denny2023conversing,wermelinger2023using}, improve programming error messages~\cite{leinonen2023using}, provide code explanations~\cite{macneil2023experiences,sarsa2022automatic}, and to provide support for students in need of help~\cite{liffiton2023codehelp,hellas2023exploring,liuTeachingCS50AI2024,hickeAITAIntelligentQuestionAnswer2023}. 

Tools, systems, and teaching practices that leverage LLMs to help students are also starting to emerge. As an example, LLMs have been integrated into code explanation tasks~\cite{denny2024explaining}, and students have been asked to write prompts that create code to solve a given problem with the intent to learn prompting~\cite{denny2024prompt}. Similarly, systems such as CodeHelp provide the possibility to paste in code and a handout, and to ask for suggestions about the code~\cite{liffiton2023codehelp}. A recent trend that is visible also outside computing education is the emergence of LLM-powered chatbots that provide the possibility to interact with an LLM~\cite{liuTeachingCS50AI2024,hickeAITAIntelligentQuestionAnswer2023,prasad2023generating,hellas2024experiences}, possibly in the context of a specific course where the chatbots are instructed to respond in a specific manner~\cite{liuTeachingCS50AI2024,hickeAITAIntelligentQuestionAnswer2023} and with additional information about the course context with techniques like retrieval-augmented generation~\cite{lewis2020retrieval}.

Although LLMs are here to stay and they bring benefits, we see that one of the issues in the recent efforts to study what sorts of tasks LLMs can be used for has been a lack of comparison to more established approaches. As an example, while LLMs can be used for responding to student queries, the same can often be done with basic search functionality. This need for comparing novel approaches with existing ones is the key motivation for our work.

We conducted a lab-based study where participants solved tasks in the context of a web software development course using a classic search bar and an LLM-powered chatbot that uses retrieval-augmented generation. The information-seeking tasks were chosen from a pool of questions that students who had previously taken the web software development course had posed to an LLM-powered chatbot. In the lab-based study, a part of the study participants were first working with the LLM-powered chatbot and then with a search bar, while a part of the study participants first worked with the search bar, and then worked with the LLM-powered chatbot. Our research questions for the present study are as follows:

\begin{enumerate}
    \item [RQ1] \rqOne
    \item [RQ2] \rqTwo
\end{enumerate}

\section{Background}
\label{chapter:background}

Large language models (LLMs) are neural network-based language models that have been trained with vast amounts of data~\cite{brown2020language}. The larger contemporary models consist of billions of parameters that are learned during model training, which is typically divided into two stages: pre-training and fine-tuning~\cite{brown2020language,radfordImprovingLanguageUnderstandinga}. The pre-training forms the basis of the model, while the fine-tuning phase is used to tailor the model for specific tasks~\cite{brown2020language,radfordImprovingLanguageUnderstandinga}.

LLMs are prompted -- given input -- that acts as the instruction based on which the LLM creates the output. LLMs can, for example, be used to translate text from one language to another, answer questions, summarize text, and generate text~\cite{devlinBERTPretrainingDeep2019,brown2020language,radfordImprovingLanguageUnderstandinga,radford2019language}. LLMs have also been trained for specific tasks and applications, such as working with source code~\cite{chenEvaluatingLargeLanguage2021}, and they have been also used to solve and create programming exercises~\cite{sarsa2022automatic,denny2023conversing}. 

As the training of large language models requires a large dataset and as the training of models is expensive, creating new models for each individual context or whenever new datasets are released is not feasible. One technique that has been used to address the issue is retrieval-augmented generation (RAG), which makes use of external data sources -- usually a vector database -- that are queried for additional information related to the user prompt before sending the user prompt with additional found information to the LLM~\cite{lewis2020retrieval}. With RAG, search results can be traced back to the original source, providing the user with the possibility to validate and trace the source of the LLM answers -- RAG has also been found to reduce hallucination~\cite{lewis2020retrieval}\footnote{We note that there is a range of fine-tuning and optimization techniques for LLMs that influence the LLM utility. We omit these for brevity.}.

Within the context of Computing Education, LLMs have recently gained plenty of attention~\cite{denny2024computing,prather2023robots}, where they have been explored for a range of tasks, including analyzing student work and creating learning resources~\cite{prather2023robots}. Importantly, for the context of the present work, their use as AI assistants have also gained attention. As an example, the AI-TA system~\cite{hickeAITAIntelligentQuestionAnswer2023} automatically answers questions from students, in order to reduce human labour. In the evaluations of AI-TA, system was fine-tuned on a dataset consisting of Piazza questions and it used RAG for retrieving data from the course materials~\cite{hickeAITAIntelligentQuestionAnswer2023}. Similarly, the ``CS50.ai''~\cite{liuTeachingCS50AI2024} is an AI chatbot that mimics a rubber duck and uses RAG for retrieving content from materials (lecture captions). In general, in evaluations, students see the AI chatbots as largely helpful~\cite{hickeAITAIntelligentQuestionAnswer2023,liuTeachingCS50AI2024} -- LLM-based AI chatbots have been deployed in other contexts as well~\cite{dam2024complete}.

The key gap in this line of research focusing on LLMs is that the contemporary LLM-based systems are typically not evaluated against (or with) other systems that serve a similar purpose. In terms of classic computing education research work, many LLM-papers could be categorized as a combination of ``Tools'' and ``Marco Polo'' -papers~\cite{valentine2004cs} that focus describing experiences from applying LLMs for a particular task. As Guzdial has pointed out, this is not necessarily a problem, as ``Marco Polo'' papers are a natural beginning of research projects~\cite{guzdial2013exploring}. There is a need, however, for investigations where the explorations involve other types of tools as well to avoid a pitfall where novel technologies are recommended even though traditional technologies might actually be better suited for the task. 

In our case, we are studying how classic search functionality fares when compared against retrieval-augmented generation large language model search. 

\section{Methodology}

\subsection{Experiment Platform}
\label{implementation:rag-chatbot}

For the present study, we adopted an existing in-house course platform with an LLM-based chatbot (without RAG-functionality). We augmented the platform with search functionality and added RAG functionality for the LLM-based chatbot. For the search functionality, we used Pagefind\footnote{\url{https://pagefind.app/}}, which is a search library that indexes given static resources and provides a JavaScript API for querying the index. For the LLM and RAG functionality, we used OpenAI's \texttt{gpt-3.5-turbo}\footnote{The \texttt{gpt-3.5-turbo} was chosen over other OpenAI models based on cost estimates formed from the prior use of the LLM-based chatbot on the course platform as well as it's similarity to recent open-source LLMs in terms of performance.} as the model and \texttt{text-embdding-3-large} as the embedding. As the vector database, we used Weaviate\footnote{\url{https://weaviate.io/}}. 

\begin{figure}[!htb]
    \centering
\begin{minipage}{0.55\textwidth}
\centering
        \includegraphics[scale=0.55]{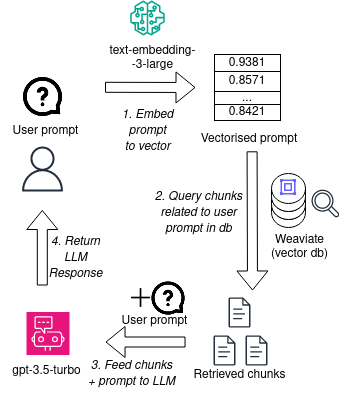}
        \caption{Flow of a user prompt when using the LLM-based chatbot.}
        \label{fig:prompt-flow}
    \end{minipage}%
\begin{minipage}{0.44\textwidth}
\centering
\begin{lstlisting}[label=listing-system-prompt, caption=System prompt of used assistant]
You are a course assistant of the course 
Web software development (WSD), answering 
students' questions. Any documents given to 
you in the prompt are materials from the 
course. When somebody asks a question about 
code, it's possible but not ensured that the 
code is given in the context. Keep the 
response 1 or 2 sentences long. Always answer 
the question, even if the context isn't 
useful.
\end{lstlisting}
\end{minipage}
\end{figure}

Whenever the LLM-based chatbot was queried, relevant information was first retrieved from the vector database, which was then added as additional information to the actual query. The system prompt was iterated over multiple times to produce meaningful but concise answers. The flow of a query is shown in Figure~\ref{fig:prompt-flow} and the final prompt used is shown in Listing~\ref{listing-system-prompt}.

\begin{figure}[h!b!]
    \centering
    \includegraphics[scale=0.45]{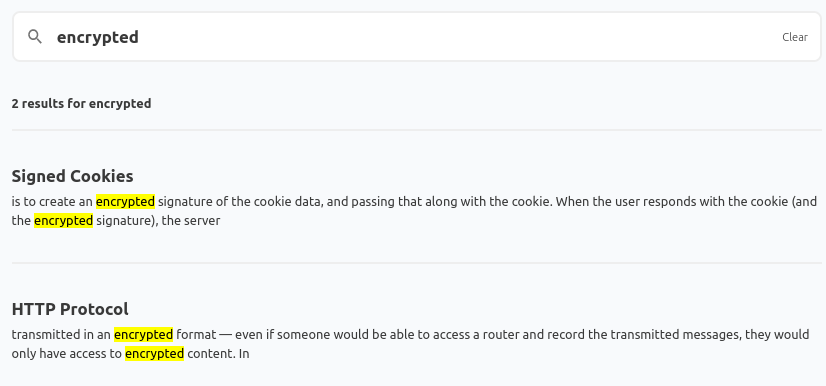}
    \caption{View of search results when typing "encrypted".}
    \label{fig:implementation-search}
\end{figure}

The search functionality was embedded at the top of the course materials as a search box which, when the user typed in text, showed a list of results with snippets of relevant text and links to the relevant pages. Figure~\ref{fig:implementation-search} highlights what the query results would look like for the query ``encrypted''. As can be seen from the figure, two search results are returned from the pages ``Signed Cookies'' and ``HTTP Protocol''. All occurrences of the searched keyword are highlighted. Clicking on any of the links would redirect the user to the corresponding course page.

\begin{figure}
    \includegraphics[scale=0.35,trim={0.8cm 7cm 1.3cm 5.2cm},clip]{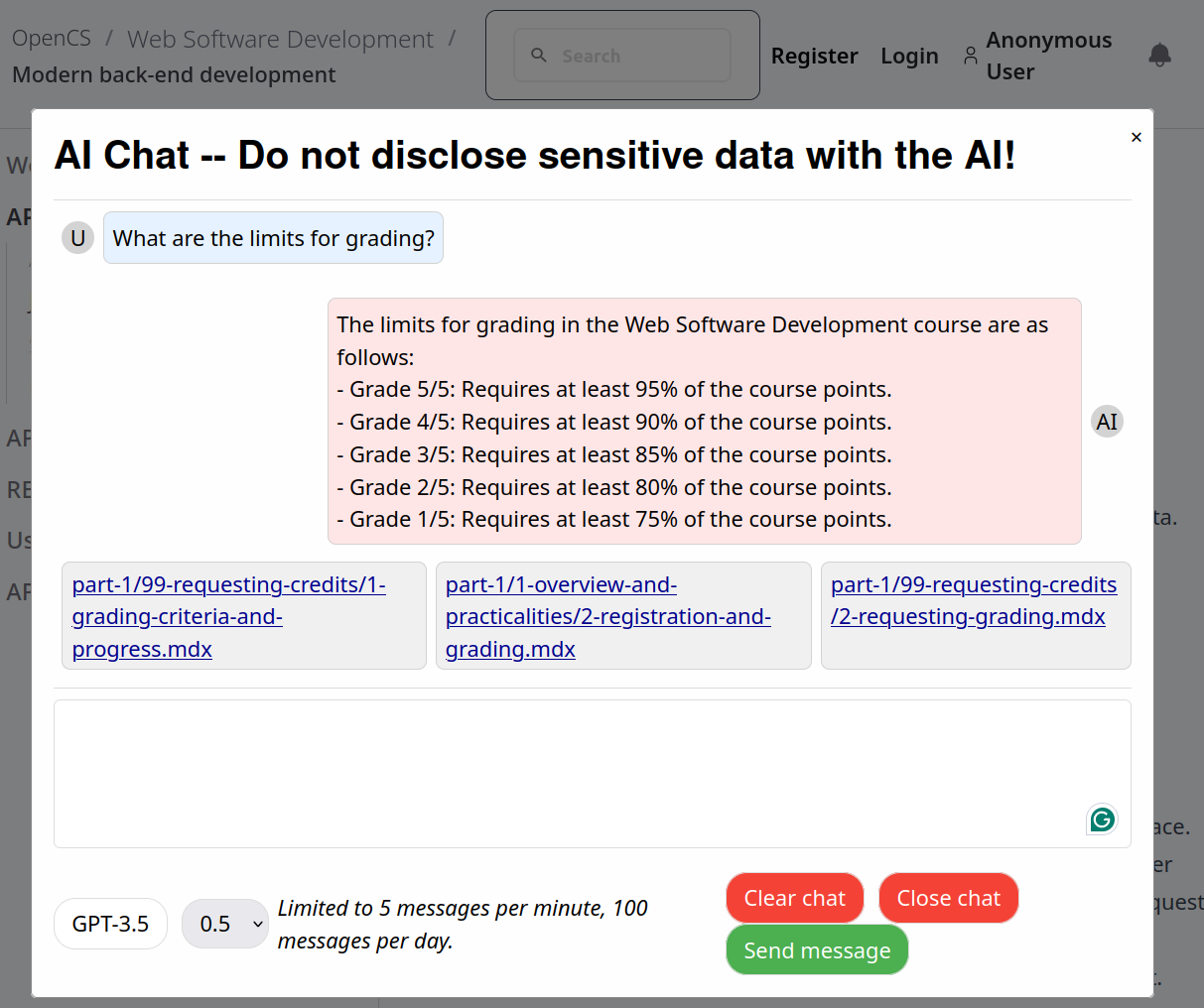}
    \caption{Chatbot dialogue with a user query and the chatbot response with links to relevant resources. The user asked what the limits for grading are, to which the chatbot described the grading limits. Clicking on any of the links redirects the user to the corresponding course page.}
    \label{fig:implementation-new-course}
\end{figure}

The RAG functionality was embedded in the existing LLM-based chatbot. 
Whenever the user typed in a query, the chatbot provided a response and potentially linked to material pages that were relevant to the query. Figure~\ref{fig:implementation-new-course} shows the response to a user query ``What are the limits for grading?''.

\subsection{Study Design}
\label{sec:3usability}

For the evaluation of the chatbot and the search functionality, we designed a lab-based summative usability test~\cite{lazarResearchMethodsHumanComputer2017} where participants complete information retrieval tasks on the experiment platform. The tasks were drawn from a database consisting of prior student interactions with the LLM-based chatbot on the course platform in the context of a Web Software Development course, which meant that they represented tasks that students had earlier tried to solve with the help of the LLM-based chatbot.
All of the tasks that were included to the study were such that they could be completed with both the LLM-based chatbot and the search functionality -- the tasks involved information seeking, explaining concepts and code, and providing code for specific tasks.

As there were two functionalities to evaluate, the chatbot and the search functionality, we designed two sets of five tasks that each participant would complete. One half of the study participants would complete a set of tasks first with the chatbot, and then another set of tasks with the search functionality, while the other half of the study participants would first complete a set of tasks with the search functionality and then another set with the chatbot. This division was created to mitigate potential biases related to the order in which the functionalities were presented (e.g., primacy and recency effect~\cite{steiner1989immediate}). We took precautions to have a similar task difficulty and similar task types in both sets, but the tasks were not identical. The tasks were completed using the think-aloud protocol, where participants were instructed to think out loud their decisions.

The study design was piloted with two Computer science students to assess the clarity of the tasks and to verify that the experiment platform was functioning as expected. The pilot revealed that in one of the tasks, the chatbot would provide a wrong answer even when the additional context provided to the LLM included the correct information. The task was changed and re-evaluated for the final study.

\subsection{Data Collection}

We advertised the study on the University course chats and by word of mouth, gaining a total of 14 participants. The participants were mainly 3rd-year Bachelor Computer Science students and 1st-year Master Computer Science students. The participants were already familiar with the course platform that the experiment platform was based on. The lab studies were conducted as one-on-one sessions, where one researcher was present with a participant, guiding the participant through the study protocol, starting with outlining the purpose of the study and asking for informed consent, and then presenting the experiment platform and the first set of tasks. Each session was screen-recorded with audio.

After a participant had completed a set of tasks with one of the functionalities, they were given a survey that had questions about the functionality that they had used, and the researcher could ask for additional questions related to the participants' answers. Then, the participant would start working on another set of tasks with the other functionality, and would at the end again be given a survey about the functionality that they had used. The survey included the following questions. 

\begin{enumerate}
    \item How did you perceive the utility of the search bar / chatbot when searching for information?
    \item How did you perceive the utility of the search bar / chatbot in general?
    \item What did you like the best about the search / chatbot functionality? 
    \item If you could improve the search bar / chatbot to fit your needs, how would you do it?
\end{enumerate}

The first two questions were collected using Likert-like scale from 1 to 5, where 1 was the lowest score (not at all useful) and 5 the highest score (highly useful), while the latter two questions were open-ended. We also collected course-specific suggestions and, for the chatbot, asked about the factuality of the answers and about the utility of the chatbot given the existence of other tools like ChatGPT. The experiment platform also logged interactions with timestamp data, allowing reconstruction of user interactions in a post-hoc manner for additional validation. 

At the end of the session, the participants were also asked for their preference related to the functionalities, asking if they had to choose, which functionality would they prefer (search bar / chatbot / neither).

\subsection{Data Analysis}

In the data analysis, we focus on descriptive statistics and do not conduct statistical tests due to the number of participants being lower than recommended for statistical testing.

To answer RQ1, \textit{\rqOne}, we provide summary statistics of the Likert-like responses to the questions posed to the participants at the end of the tasks sets, accounting for the order in which the functionalities were presented to the participants. We further provide summary statistics related to user preferences of the functionalities, and outline key observations of what the participants liked the best about the search and the chatbot and how they would improve the functionality to provide additional information related to the perceived utility.

To answer RQ2, \textit{\rqTwo}, studying the screen-recordings and the interaction data, we manually graded the user responses to the tasks, giving each task response 0, 0.5, or 1 point depending on whether
they were false, somewhat correct or fully correct, providing summary statistics of the correctness of the tasks. In addition, using the timestamps from interaction data, we measured the time that each of the tasks took to complete, providing the time information as additional data related to performance.

\subsection{Ethics Statement}
The study was based on informed consent. Based on the national regulations of the study context, no ethics review was required. The student interactions with the LLM-based chatbot that were used to form the tasks were from students who had consented to sharing their data for research purposes. The research team had access only to the prompts, but had no information of who had written the prompts.

\section{Results}

\subsection{Usefulness and Preference}

The first objective of the user study was to study the utility of the LLM-based AI chatbot and the search functionality.
Group A first
used the search bar and then the chatbot, while group B first used the chatbot
and afterwards the search bar. The summary statistics of the Likert-like responses are shown in Table~\ref{table:values}. 

Overall, averaged over both populations, the utility of searching for information with the chatbot was rated 4.57/5 while the utility of searching for information with search was rated 4.36/5. The averaged general utility of the chatbot was 4.43/5 while the averaged general utility of the search was 4.21/5. In both cases, the chatbot was perceived slightly more useful.

For the group B that started with the chatbot, the median utility of both the chatbot and the search functionality was 5/5, while for the group A that started with the search functionality, the median utility of both the chatbot and the search functionality was 4/5. In particular, the general utility of the search functionality was viewed as relatively low by the group starting with the search functionality (3.71/5). The contrast is rather stark when comparing the result with the group A that started with the chatbot, and who rated the search functionality very high (4.71/5) after first working with the chatbot.  

\begin{table*}[h!]
\caption{Summary of participant perception in user survey. The columns in orange point to population A, and the columns in green point to population B.  The values are on a Likert-like scale, ranging from 1 to 5 (1 = not at all useful, 5 = very useful for rows 1, 2 and 4; 1 = not at all correct, 5 = to a great extent correct for row 3). The symbol $\mu$ denotes mean, $m$ denotes median.}
\label{table:values}
\begin{tabular}{l|l|
>{\columncolor[HTML]{9AFF99}}l 
>{\columncolor[HTML]{9AFF99}}l |
>{\columncolor[HTML]{FFCE93}}l 
>{\columncolor[HTML]{FFCE93}}l |
>{\columncolor[HTML]{FFCE93}}l 
>{\columncolor[HTML]{FFCE93}}l |
>{\columncolor[HTML]{9AFF99}}l 
>{\columncolor[HTML]{9AFF99}}l |ll}
 &
   &
  \multicolumn{2}{l|}{\cellcolor[HTML]{9AFF99}Chat-first} &
  \multicolumn{2}{l|}{\cellcolor[HTML]{FFCE93}Search-first} &
  \multicolumn{2}{l|}{\cellcolor[HTML]{FFCE93}Chat-second} &
  \multicolumn{2}{l|}{\cellcolor[HTML]{9AFF99}Search-second} &
  Chat Both &
  Search Both \\ \hline
 &                                                                         & $\mu$ & $m$ & $\mu$ & $m$ & $\mu$ & $m$ & $\mu$ & $m$ & $\mu$ & $\mu$ \\
 & \begin{tabular}[c]{@{}l@{}}General utility\end{tabular}  & 4.43  & 5    & 3.71  & 4    & 4.43  & 4    & 4.71  & 5    & 4.43  & 4.21  \\
 & \begin{tabular}[c]{@{}l@{}}Search utility\end{tabular}   & 4.71  & 5    & 4.14  & 4    & 4.43  & 4    & 4.57  & 5    & 4.57  & 4.36  \\
\end{tabular}
\end{table*}

\begin{table}[]
\caption{Preference toward functionalities across the two groups. }
\label{tab:preference}
\begin{tabular}{l|lll}
 & Chat & Search & Neither \\ \hline
Chat-first group & 0 & 6 & 1 \\
Search-first group & 5 & 0 & 2
\end{tabular}
\end{table}

At the end of the study, the participants also provided information of which functionality they would prefer. The statistics of the preferences are shown in Table~\ref{tab:preference}. We observe a clear split between the groups -- interestingly, both groups preferred the functionality used in the second part (or chose neither). The group A that started with the search mainly chose the chatbot as their prefererred choice, while the group B that started with the chatbot mainly chose the search as their preferred choice.

Finally, to provide additional information related to the utility and preference, we studied the open-ended answers that the participants gave when prompted for what they liked the best about the functionalities and how they would improve the functionalities. Overall, when considering the search functionality, it was liked for its responsiveness and quickness, with no need to hit `enter' to get search results, and search results generating without having to type all letters of a keyword. Further, it was appreciated that the search results came with a part of the result document, highlighting queried keywords. At the same time, some participants noted that the search returned too many irrelevant results, and that the order of results should be improved. One participant also noted that the search should account for synonyms and consider the context, e.g., by using Elasticsearch's Vector search~\cite{ElasticSearchWhatVector}.

When considering the chatbot, many participants mentioned that they enjoyed that the chatbot often showed related documents in addition to the results\footnote{While carrying out the study, we also noticed that the participants often visited the related source documents to verify that the chatbot response is correct.}. In addition, most participants appreciated that the chatbot summarized the lengthy parts of relevant course documents into condensed information. When considering improvements, some noted that the chatbot did not always recommend sources and that sometimes the recommended sources were irrelevant. Some noted that the response took quite a bit of time (often multiple seconds). One participant who had started in the search group noted that the chatbot should also show parts of the text of the retrieved materials similar to the search functionality, as this could speed up verification. Another participant voiced out dislike towards the chatbot, mentioning that it could encourage laziness and increase the reliance on assistants and tools.

\subsection{Task Performance}

The second objective of the study was to assess the performance of the participants in the tasks, comparing the two functionalities. The performance was assessed using two metrics: the manually graded correctness of the answers and the time to complete each task.

\begin{figure*}
    \includegraphics[scale=0.4, trim={4cm 0 4cm 0},clip]{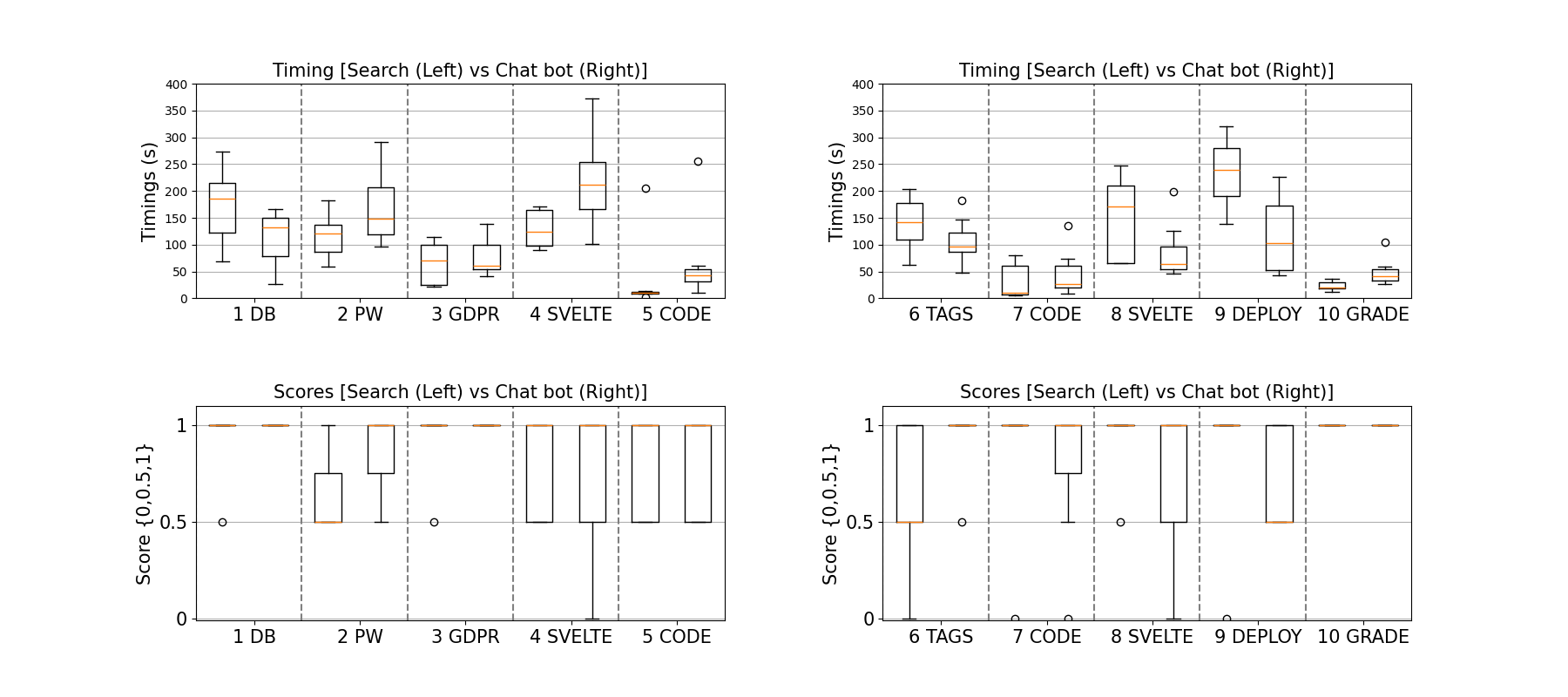}
    \caption{Comparison of performance, broken down by the time to complete an exercise and the received score, grouped by exercise. For every exercise, the left box depicts the search bar and the right box the chatbot.}
    \label{fig:box-performance}
\end{figure*}

Figure~\ref{fig:box-performance} highlights the task completion time
and scores over the individual tasks.
On the x-axis, we have the tasks, broken down by
groups A and B. On the y-axis, we have the task completion time
in seconds, and the score of an answer, respectively.
It is vital to note that for all exercises shown in the plot, there are
two boxes: the first one always refers to the search bar and the right
one always refers to the chatbot. This means, that in the plots on the
left column, the boxes of group A are always the left ones, while in the plots
of the right column, the boxes of group A are always the right ones.

When looking at the boxes for each exercise, we can not directly deduce
that one functionality helps to complete the tasks faster or leads to a higher score. Therefore, we analysed the data for each question, looking at the correctness and completion time of individual answers, and highlighting differences.
For instance, Q10 shows a good example,
in which the search bar was superior. The question requires the 
user to look up the grading scheme and report
what grade they would get as they have a score of 86\%. We can see that the
completion time was consistently higher with the chatbot ($\mu=49.00s$) than with the
search bar ($\mu=23.43s$). Similarly, we observed that the Q4 was faster to complete with the search than with the chatbot -- here, the question related to how state management works with Svelte. It seemed that many participants using the chatbot did not understand the chatbot response. As a result, this subset of participants consulted the related sources and manually searched for relevant information.

We had expected to see a similar outcome for Q3, in which the participants
were prompted to look up and report the fines for violating the GDPR regulations.
However, in this case, the related page was longer and richer in information, and some
participants struggled to scan and find the required information to report.

Next, we would like to discuss the task completion times for Q8 and Q9.
Both tasks were quite complex, in that the users had to summarize a large
chunk of information, spread across a long course page. Unsurprisingly
the participants finished the task particularly quickly with the help of the chatbot,
as the chatbot could summarize the contents of the page in a few concise
paragraphs. The mean values for Q8 and Q9 for the chatbot are $\mu=87.29s$ and $\mu=117.57s$, respectively, and $\mu=147.71s$ and $\mu=234.14s$ for the search bar. On the other hand, when we look at the scores of Q8 and Q9, we see that the chatbot scores were lower than with the search bar counterpart. When inspecting the answers we noticed that for those two questions, it happened frequently that no related source documents were found, or that the wrong documents were found -- the prompts were often such that they could be improved considerably. In those cases, the answer to the question was then based on hallucinated information, which the participants often relied on. 

Next, we noticed that the scores for Q2 -- listing the practices to
store passwords securely -- were signifiantly better, when
participants had access to the chatbot. The reason here, is that with the search functionality, the participants sometimes reported the first security guidelines that they found and omitted the following ones. With the chatbot instead,
the participants got access to all requested security guidelines immediately. 
This points to one negative symptom that we observed in some 
participants working with the search bar: some participants would search for
the relevant course page and not read through the entire course page,
but instead only look for the highlighted section. 
In this case, all
participants saw that they should hash their passwords, but only 2 out of 7
participants reported that the password should be salted. 

\section{Discussion}

When looking at the assistants isolated, it is straightforward to say
that the participants perceived both functionalities useful. The general utility of both assistants was well
appreciated, and the participants found aspects that they liked in both of the functionalities. On average, the participants ranked the utility of the chatbot somewhat higher than the utility of the search functionality. There are, however, studies that highlight a placebo effect that has been associated with AI systems~\cite{kloft2024ai}, where users of AI assistants believe that the assistants will help them perform better even if that would not be the case. 

When considering participants' preference between the two functionalities, there was a clear split between the two groups, where the groups always preferred the latter functionality. That is, the group that started with the chatbot preferred the search and vice-versa. It is possible that this stems partially from the recency effect~\cite{steiner1989immediate}, where the items that come later are remembered better. It is also possible that this relates to the participants ``learning the ropes'' when working on the first set of tasks. We do note that the questions were explicitly selected (and tested) so that the chatbot could be of assistance and the chatbot could be used to solve them. 
Despite this, we observed that some participants received hallucinations in the answers, mainly due to poor prompting.

When considering the performance, we can see that the two functionalities have their strengths and use cases. The chatbot was best at summarizing 
large amounts of text into a few concise paragraphs. The search bar, on the other hand, was best at quickly finding relevant pages. We see that the differences in performance were related to three key points:

\begin{enumerate}
    \item Response speed: For search, initial search results were already shown when typing text, while for the chatbot, the user had to formulate a prompt and wait for a few seconds for a result. 
    \item Completeness and correctness: For search, the users would see potentially relevant snippets in the search results, while for chatbot, the users would see summarized outputs that potentially included hallucinated information.
    \item Need for verification: For search, for some of the tasks, all relevant information was in the text snippets and there was no need for verification, while with the chatbot, users often had to double-check the results by consulting the relevant materials.
\end{enumerate}

One of the study participants also voiced concerns related to the chatbot encouraging laziness and increasing reliance on assistants and tools. None of the participants noted the same for the search functionality. We consider that this likely stems from search being a commodity, while the chatbot can still be seen as a novel tool. This leads us also to asking how would introducing these two functionalities change how participants learn, and approach exercises? Furthermore, would incorporating these two functionalities decrease the learning effect and promote over-reliance on the functionalities? We find that both functionalities have their risks and opportunities, in other words, they can promote learning when used properly, but harm learning when used incorrectly. 

While the search bar may seem hard to abuse, there is a risk that
participants would spend less time reading course materials and potentially omit relevant information. While traditional search
engines like Google provide similar functionality, in this case,
the search bar is tailored to the course materials and participants
will still navigate to the course website, possibly giving them a false
sense of security. This danger could be seen in Q2 -- finding out
about the guidelines for storing passwords -- when participants overlooked
about the need for a salt. 

Due to the power of LLMs, the chatbot has greater potential for misuse, when compared to the search bar. Participants may use the chatbot to solve assignments instead of working on them on their own, they may stop questioning the output of the chatbot, and in general, over-rely on the help of the chatbot. Prior research has suggested ways to reduce these issues, including adding guardrails that prevent participants from getting too much help with assignments and adding rate-limitations so that participants are encouraged to use the chatbot more intentionally~\cite{liuTeachingCS50AI2024,liffiton2023codehelp}. 

We acknowledge that there are a handful of limitations to our study. First, our sample size was small (n=14), and it consisted of volunteers, indicating sampling bias. Second, we used the \texttt{gpt-3.5-turbo} model, which is known to perform somewhat worse than the \texttt{gpt-4} model that recent AI assistant studies (e.g.~\cite{hickeAITAIntelligentQuestionAnswer2023,liuTeachingCS50AI2024}) have used. At the same time, the \texttt{gpt-3.5-turbo} is on par with open-source LLMs, providing information on how models that can be run locally with sufficient hardware may perform. Third, we also acknowledge that the model was not fine-tuned with existing question data, and we did not conduct any optimizations such as caching that could have improved the performance of the chatbot. Fourth, we note that the tasks for the study were hand-selected so that both the chatbot and the search functionality could be used to solve the tasks. It is possible that an in-situ evaluation would lead to different outcomes. Fifth, we also acknowledge that the study did not include tasks where students would have had to implement programs due to time limitations. We see that it is likely that the chatbot could have performed better for such tasks, at least if the tasks would not use technologies that the model would not know of~\cite{hellas2024experiences}. Finally, we acknowledge that the interfaces of the two functionalities were different and that the user experience can already influence study outcomes. The LLM-chatbot used an existing interface that the participants were already familiar with, with the exception of seeing the source documents, while the search interface was new.

\section{Conclusion}

In this study, we studied the utility, preference, and performance of an LLM-based AI chatbot with RAG functionality and a search bar in the context of web software development related information-seeking tasks. 

To summarize, when considering the perceived utility, the participants were satisfied with both the chatbot and the search. It seems like there is no universal preference towards a specific assistant, but that the utility is rather context-dependent and related to the task. Interestingly, we observed that participants who started with that chatbot mainly stated that they would prefer the search, while participants who started with the search mainly stated that they would prefer the chatbot.

When considering the performance, it seemed that the chatbot was better at tasks that required summarizing large amounts of text into a few concise paragraphs, but users needed to often verify the results. The search bar, on the other hand, was better at quickly finding relevant pages, without having any delay of the results. Additionally, the search bar did not have the introduced risk of hallucination, yet it was easier to miss small details.

Our results continue highlighting the utility of LLMs and provide further supporting evidence to earlier studies on using LLMs as support systems (e.g.~\cite{liuTeachingCS50AI2024,hickeAITAIntelligentQuestionAnswer2023}). At the same time, our results also show the utility of existing and more traditional systems, suggesting that integrating both functionalities to learning platforms could provide better outcomes than relying on only one of them. 

\balance
\bibliographystyle{ACM-Reference-Format}
\bibliography{main}

\end{document}